\documentclass[structabstract]{aa}

\usepackage[latin1]{inputenc}
\usepackage{txfonts}
\usepackage{graphicx}
\usepackage{epsfig} 
\usepackage{epstopdf}
\usepackage{natbib} 
\bibpunct{(}{)}{;}{a}{}{,}

\begin{document}
\title{Fitting of dust spectra with genetic algorithms}

\subtitle{I. Perspectives \& Limitations}

\author{A. Baier\inst{1}
\thanks{\emph{email to:} angela.baier@univie.ac.at}
\and F. Kerschbaum\inst{1}
\and T. Lebzelter\inst{1}}

\institute{University of Vienna, Institute of Astronomy; T\"urkenschanzstra\ss{}e 17, A-1180 Wien, Austria}
\date{Received 24 12 2009 / Accepted 12 03 2010}
\abstract {} {We present an automatised fitting procedure for the IR range of AGB star spectra. Furthermore we explore the possibilities and boundaries of this method.} {We combine the radiative transfer code DUSTY with the genetic algorithm PIKAIA in order to improve an existing spectral fit significantly. } {In order to test the routine we carried out a performance test by feeding an artificially generated input spectrum into the program. Indeed the routine performed as expected, so, as a more realistic test set-up, we tried to create model fits for ISO spectra of selected AGB stars. Here we were not only able to improve existing fits, but also to show that a slightly altered dust composition may give a better fit for some objects.} {The use of a genetic algorithm in order to automatise the process of fitting stellar spectra seems to be very promising. We were able to improve existing fits and further offer a quantitative method to compare different models with each other. Nevertheless this method still needs to be studied and tested in more detail. }
\maketitle
\section{Introduction}
Asymptotic Giant Branch (AGB) stars \citep[e.g.][]{agbbook} are one of the main sources of dust in the universe. According to their evolutionary state they produce different amounts and species of dust. At the onset of their mass loss stars with solar metallicity exhibit a mainly 
oxygen-rich dust mineralogy, which consists of oxides and silicates \citep[e.g.][]{posch99, jaeger03}. At later stages of
their evolution, the star's chemical composition changes depending on its mass, which may lead to a mineralogy dominated by carbon-rich dust such as amorphous carbon, carbides, sulphides and nitrides \citep[e.g.][]{hony04}.

Dust formation is of immense importance for the overall behavior of these stars. It has an effect on the dynamics of their atmospheres
and thus on their mass loss. As a consequence, the spectral appearance changes as well. The dust can produce prominent features in the infrared regime of a spectrum. Many of those dust species might be still unknown, others produce spectral features in the same wavelength region and thus are difficult to distinguish. Another aspect is the dependence of the feature shape on the dust grain shape itself. 

Up to know, the fitting of those spectra has mostly been done manually. A routine to automatise this process, which also allows to quantify the results would be a big asset in this field of research. 
In this work we are taking first steps into this direction. By combining a genetic algorithm with a widely used radiative transfer code we want to show that it is possible to veer away from the traditional way of hand fitting to a more independent procedure.   

\section{The Fitting Procedure}
\subsection{The problem}
\label{tp}
As mentioned above, the spectrum of a low mass late type star can exhibit strong dust features in the infrared range. The mostly complex dust composition of the star's shell can make the determination of its exact composition difficult.

In order to determine the detailed dust composition of a star a synthetic spectrum is fitted to an observed one. The parameters for the synthetic spectra are chosen to  include those dust species, which show features at the same wavelengths as the spectrum of the observed object. This may then eventually lead to a satisfying fit and thus to the identification of certain dust species. Since most stars do not only house a single dust species in their surroundings, their respective infrared spectrum can be quite complex, showing features that are the result of a certain blend of minerals. An unambiguous determination of the dust composition can constitute a complex challenge, since the spectral behaviour of dust does not only depend on the correct ratio of abundances, but also strongly on such parameters as the dust temperature, the optical depth and the grain shape itself \citep[see e.g.][]{posch07b}.

One of the most commonly used radiative transfer codes to produce these synthetic spectra is DUSTY \citep{ivezic97}. During the last years it has proven to be a very useful tool to gain new insights into the dust mineralogy of evolved stars, although it has to be noted that DUSTY does not treat the dust opacity as a function of distance from the star. The assumption of an opacity $\kappa$ which is invariant to the stars radius is, however, unrealistic.

Nevertheless DUSTY is a quite powerful tool and serves as a very good starting point in means of determination of dust composition  for highly evolved stars. Although this method of fitting an IR dust spectrum of a star is a robust one, it holds a downside which always needs to be taken into account when looking at the results. Casually speaking, the problem is the scientist himself. First, in order to produce a ``good'' fit a certain amount of experience and intuition is definitely an asset. Second, this experience can also be a drawback, since different people may produce equally good fits with a completely different set of parameters. This raises the need to quantify those models in order to compare them with each other. Furthermore a process which automatises the fitting procedure, in terms of finding the correct parameter values of the desired fit and thus produces a result which is less biased by human intervention, would serve the purpose of quantification even more.
 
Inspired by a preliminary study by \citet{dijk07} an attempt has been made to combine DUSTY with a genetic algorithm. Following Dijkstra's example the  publicly available genetic algorithm PIKAIA \citep{charb95} has been used for this task. 

\subsection{DUSTY in a nutshell}
DUSTY is used to perform radiative transfer calculations for objects of different nature. The code is designed to deal with the radiation from some source which is modified by a dusty region. In our  case this is a central star wrapped in a dust shell. Since the light emitted by the star is scattered, absorbed and reemitted by the dust shell, the spectrum emerging is usually the only way to gain information about the dust and the object hidden inside.

DUSTY offers users an integral equation for the spectral energy distribution introduced by \citet{ivezic97} in order to solve the radiative transfer problem. Beside the basic free parameters such as the stellar temperature, the chemical composition, the grain size distribution, the optical depth and many more, it also allows to chose between a spherical or a plane-parallel geometry. 
\subsection{Genetic algorithms \& PIKAIA}
\begin{figure}[h]
\centering
\includegraphics[bb=0 0 316 515,scale=0.5]{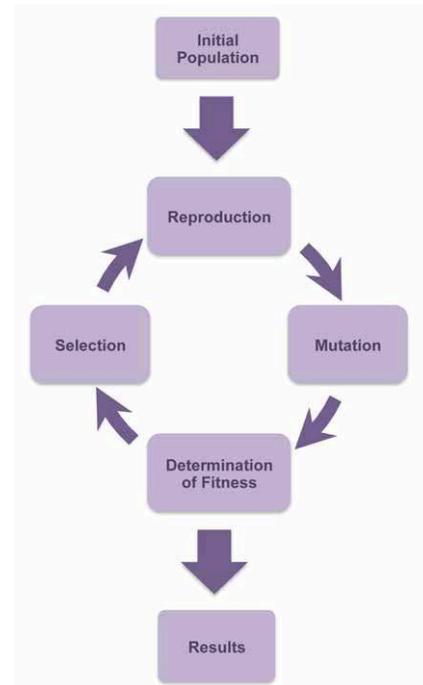}
\caption{Basic concept of a genetic algorithm}
\label{genal}
\end{figure}
Genetic Algorithms (GAs) are searching techniques used in computation for optimisation of a given problem. To assure a healthy and steady evolution a certain size of a population is required as well as the offspring showing a certain range of ``fitness'' to secure the variability of the process.

With these considerations in mind a genetic algorithm can be defined as the implementation of the evolutionary principle in a computational context.  A simplified flowchart is shown in Fig.\ref{genal}. This basic version gradually improves via successive iteration, but does not maximise in a strictly mathematical point of view. Most on genetic algorithms based codes comprise further strategies to face this problem.

The genetic algorithm-based optimisation routine called PIKAIA \citep{charb95} is written in a very user-friendly way. Thus, it is primarily used as a learning tool rather than a science tool. Still, this ``easy-to-use'' policy makes it a very attractive choice for optimisation problems of relatively low complexity. Nevertheless, one always has to keep in mind that the performance efficiency is often sacrificed to the clarity of this code.

PIKAIA maximises a problem over a fixed population and a given number generations. This means that PIKAIA does not strive to optimise a population until a certain criterion is fulfilled, but rather until a given number of generations is reached.

The code is written in FORTRAN and sticks to the conventions of ANSI-FORTRAN 77.  Although PIKAIA is also available in JAVA or as a parallelised version, the decision to use its original FORTRAN version is quite obvious since DUSTY is also written in the same language and thus a certain kind of code persistency is preserved.

Genetic Algorithms are used to solve manifold problems. In astronomy itself GAs are a common tool for optimisation.\citet{hetem07} make use of a GA in order to improve already existing models of protoplanetary disks. In his work on the transition region from the chromosphere to the corona of the sun \cite{peter01} used PIKAIA to optimise his Gaussian fits of various lines. \citet{noyes97} used the code for optimisation of the derived orbital parameters in their work on a possible planet orbiting $\rho$~CrB.  In a much larger context genetic algorithms are also used for example to model the interaction of galaxies \citep{theis98, theis01}.
\citet{metcalfe00} as well as \citet{mokiem05} used  the optimisation code PIKAIA as an improvement and automatisation of the fitting routines of the respective studied objects. This is in the case of the former, the observed pulsation frequency of the white dwarf GD 358 and in the case of the latter the combination of the non-LTE stellar atmosphere code FASTWIND \citep{puls05} with PIKAIA for the spectral analysis of early-type stars. 
Finally, \citet{canto09} used an, as they call it, asexual genetic algorithm for optimization and model fitting of various problems. In their approach they apply an asexual reproduction scheme, which means the offspring are only produced by one parent -- not unlike bacteria -- and not by two parents as used in a traditional GA. As possible applications for this method a typical optimization problem as well as the fitting of the spectral energy distribution for a young stellar objects is presented.

\subsection{PIKAIA \& DUSTY -- a convenience marriage }

\begin{table}[th]
\centering 
\caption{Dust species used for fitting the ISO spectra with DUSTY} 
\begin{tabular}{c c c c } 
\hline\hline 
f-Number & Species & Formula &  Ref\\ 
\hline 
$f_1$& Amorphous aluminum oxide& Al$_2$O$_3$                                  &  1\\
$f_2$ & astronomical silicate  &                          SiO$_4$                             &  2\\
$f_3$ & Amorphous pyroxene (100K)& Mg$_x$Fe$_{1-x}$SiO$_3$        &  3\\
$f_4$ & Amorphous melilite                      & Ca$_2$Al$_2$SiO$_7$              & 4\\
$f_5$ & Amorphous olivine                                        & Mg$_{0.1}$Fe$_{1.2}$SiO$_4$ & 5,6\\
$f_6$& Crystalline magnesiowustite                      & Mg$_{0.1}$Fe$_{0.9}$O              & 7\\
$f_7$ & Amorphous Mg-Fe-silicate      & MgFeSiO$_4$                                & 8\\
$f_8$ & Crystalline spinel                                        & MgAl$_2$O$_4$                       & 9\\
\hline
\end{tabular}
(1) \citet{begemann97}; (2) \citet{ossenkopf92};  (3) \citet{henning97};  (4) \citet{mutschke98}; (5) \citet{jaeger94}; (6) \citet{dorschner95}; (7) \citet{henning95}; (8) \citet{hanner88}; (9) \citet{posch99}
\label{tbldustspec} 
\end{table}
As already mentioned in Sect.~\ref{tp} fitting of dust spectra of AGB stars without the help of a suitable routine can be challenging and may lead to ambiguous results. In our approach, the DUSTY code was connected with the genetic optimisation algorithm PIKAIA  which allowed to produce  spectra, that are less dependent on the individual experience of the respective scientist \citep[see][]{dijk07}. Being based on the evolutionary natural selection process, PIKAIA tries to maximize the function $g(\lambda)=[F(\lambda)-F_m(\lambda)]^{-2}$, where $F(\lambda)$ is the observed spectrum and $F_m(\lambda)$ represents one model spectrum calculated with DUSTY.
 
In order to optimize $g(\lambda)$, PIKAIA creates an initial population of trial solutions for the DUSTY models $F_m(\lambda)$ by using random values for the input parameters. Starting from there, pairs of trial solutions are selected which are then used to bred two new solutions, i.e. the offspring. This way a new population with the same number of trial solutions is created, and then, the breeding process starts again until the final generation is reached.

As input parameters, for DUSTY the following have to be chosen before the calculation is started and will not be changed by PIKAIA. DUSTY's wavelength grid has been kept down to only 105 grid points in order to keep the runtime for each model as small as possible considering the big overall number of models to be calculated. 

\begin{figure}[h]
\centering
\includegraphics[bb=15 9 576 783, scale=0.55]{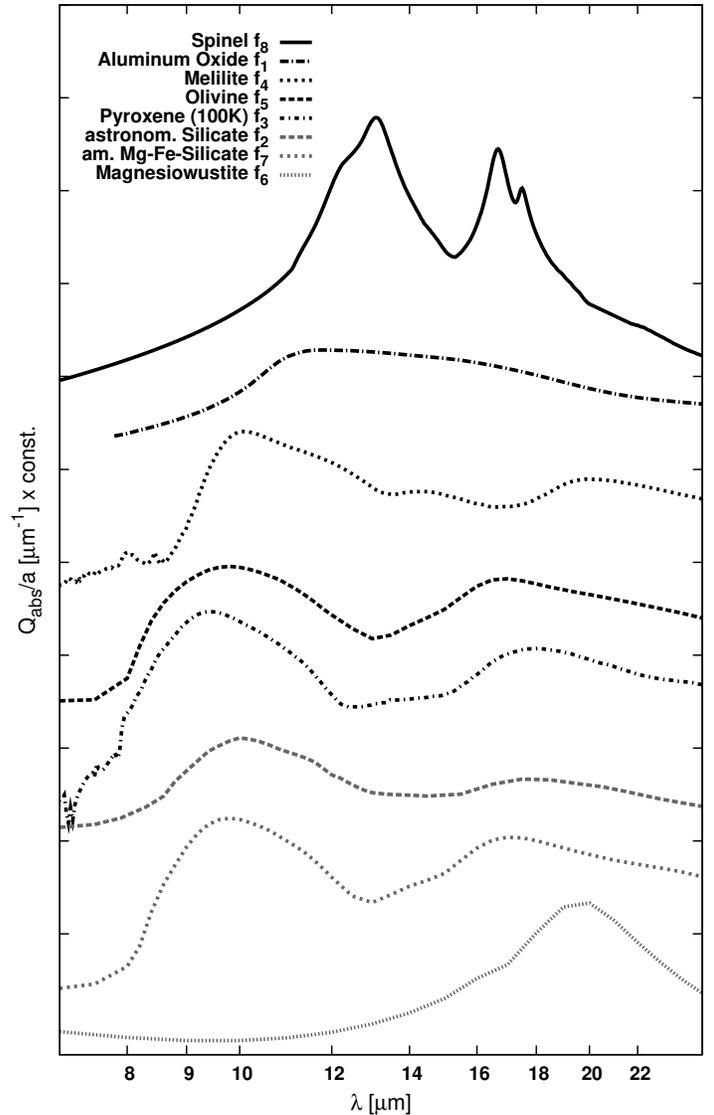}
\caption{Dust absorption data of the dust species used for fitting the ISO spectra with DUSTY}
\label{dustspecies}
\end{figure}
For the grain size distribution MRN as well as for the ratio between the outer shell radius $R_{out}$ and the inner shell radius $R_{in}$, the default settings have been kept. This grain size distribution, following \citet{mathis77} is given by a power law of the following form
\begin{equation}
n(a)\propto a^{-q} \quad \mathrm{ for } \quad a_{min} \leq a \leq a_{max} 
\end{equation}
with $n(a)$ being the number density of the dust grains with a radius a, and with the default parameters $q = 3.5$, $a_{min} = 0.005\,\mu$m and $a_{max} = 0.25\,\mu$m. For $R_{out}/R_{in}$ a value of 1000 was chosen, since any ratio above 100 has only a very small influence on the shape of the MIR spectrum. 

The effective stellar photospheric temperature $T_*$ has also to be chosen beforehand. DUSTY then approximates the photospheric spectrum by a single black body. For this first stage of modelling, it was also decided that the dust temperature $T_{dust}$ and the optical depth $\tau$ are kept throughout the calculation process. Both values have a big influence on the spectral shape and thus should be known to a certain extent already beforehand. 

Finally the pool of dust species used by DUSTY is also fixed during the whole run. Throughout this paper the dust species will be referred to as $f_1$ to $f_8$. The key to this numbering scheme can be found in Table~\ref{tbldustspec} as well as a graphical representation of the optical constants in Fig.~\ref{dustspecies} where the main features in the studied wavelength range can be seen. Their respective abundances are subject of PIKAIA's optimisation routine. In order to avoid completely unrealistic combinations of abundances, the values have always been set to a range that seemed most likely for the respective object. 

For PIKAIA two of the most important input values are the generation and population number. If these values are too small the genetic algorithm cannot act as intended, which means, values sufficiently high for the treated problem must be chosen. For our experiments we used as pop/gen values either 90/100 or 100/200, depending on the complexity of the observed spectrum. 

PIKAIA qualifies each fit using a user supplied fitness function. Here we used the straight forward $\chi{}^2$ assuming that
\begin{equation}
\mathrm{fitness}=\frac{1}{\chi{}^2}.
\end{equation}
For each model the respective $\chi{}^2$ has to be calculated and then used to generate the fitness value which is used for the internal ranking of each model. At the end of the evolutional process, PIKAIA provides the respective parameter values together with the fitness value as an output for the user. This allows us to compare different fits in a quantitative way.
\begin{figure*}
\centering
\includegraphics[bb = 26 51 560 267,scale=1.0]{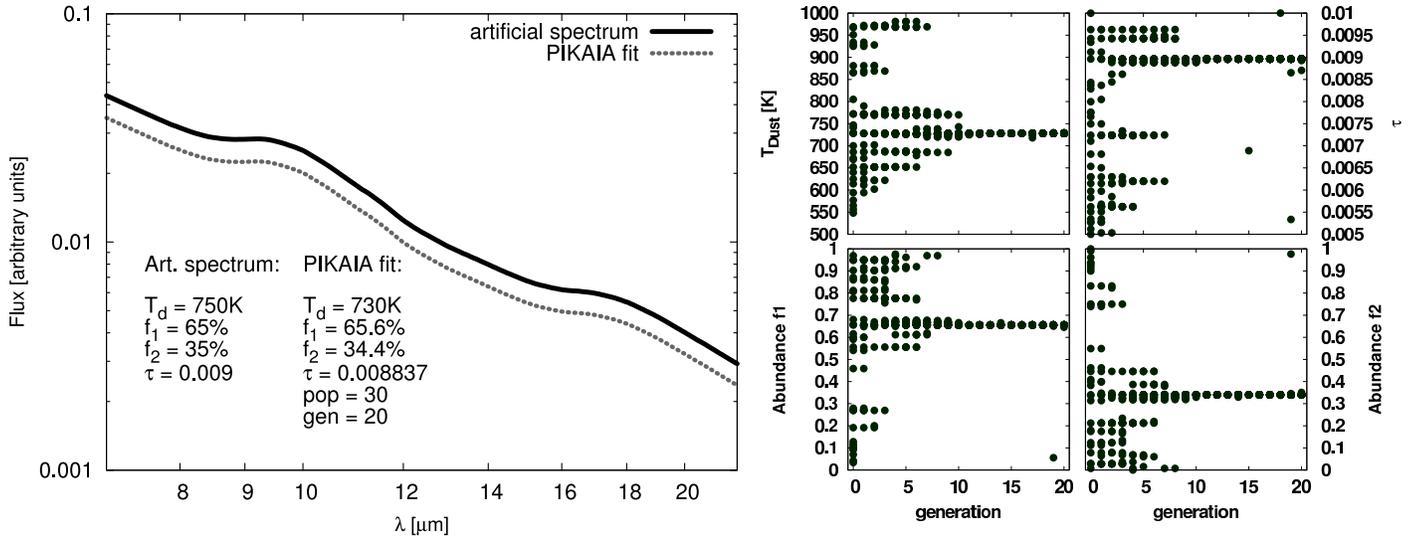}
\caption{On the left panel the result of fit number 7 to the artificial spectrum can be seen. This is the best fit which could be achieved with the highest number of free parameters. The values for the artificial input spectrum are shown in comparison with the results of the fit. Check also Table~\ref{perres} for the results of the other test runs and the associated fitness values. On the right panel the distribution of the parameters for each generation is shown. It can be clearly seen, that the more advanced we are in the evolutional process the more certain values start to dominate each population and finally converge to a final result. Of course no population consists only of 100\% perfect individuals. This also explains the occasional outliers even further in the process.}
\label{dvd}
\end{figure*}

\begin{table*}[th]
\centering 
\caption{Performance test results. The fitness has been normalised to the fittest model (nr. 2) in this test.} 
\begin{tabular}{c c c c c c c c c} 
\hline\hline 
\# & pop & gen & $T_{dust}$ &  $f_2$ & $f_3$ & $\tau_{10\mu{}m}$ & free paras& fitness\\ 
\hline 
0 & artificial& spectrum& 750 & 65 & 35 & 0.009 & - & - \\ 
\hline 
1 & 10 & 5 & 754 & - & - & - & 1 & 0.019\\
2 & 10 & 5 & - & 65 & 35 & - & 1 & 1.000\\
3 & 10 & 5 & - & - & - & 0.00898 & 1 & 0.023\\
4 & 10 & 5 & - & 66.5 & 33.5 & 0.00878 & 2 & 0.552\\
5 & 10 & 5 & 752 & 50.3 & 49.7 & 0.00827 & 3 & 0.664\\
6 & 30 & 20 & - & 66.6 & 33.4 & 0.00783 & 2 & 0.777\\
7 & 30 & 20 & 730 & 65.6 & 34.4 & 0.008837 & 3 & 0.969\\
\hline
\end{tabular}
\label{perres} 
\end{table*}
\section{Performance Test}

In order to investigate the performance and the overall functionality of the two combined programs a simple test setup has been designed. For this task an artificial spectrum has been created with DUSTY. Following the plan to keep the fake input spectrum as simple as possible, the stellar parameters have been chosen to represent a more or less average mira star with a silicate only dust composition. 

This artificial spectrum has then been fed into PIKAIA as the ``observed'' spectrum. For this performance test two combinations of generation and population numbers have been used. Furthermore, for each of the test runs  some of the parameters have been kept with a fixed value while others where allowed to be chosen freely. The value used to qualify the fits is the already introduced ``fitness''. As can be seen in Table~\ref{perres}, it does not only depend on the number of free parameters but also on which of the parameters have been left free.  For calculations  with  a very low number  of population members and generations, our procedure obviously has the least difficulties when all parameters except the dust composition are given in advance.  Still it has to be pointed out that even for the low gen/pop combination the best result has been achieved with all parameters left free. This shows that the genetic algorithm operates most efficiently when given a sufficient amount of freedom concerning the parameters. Nevertheless, when comparing the fitness values of fits 5 and 7, it is also shown very clearly that in the case of no fixed parameters, a low number of populations and generations -- such as in the case of fit 5 -- is not sufficient anymore to retrieve a reasonable result. 

The generation and population numbers may appear to be very low considering that one would expect both values to be higher to receive acceptable results from the algorithm. This is indeed true, when the code is applied to real observed spectra. Still for a simple test setup like the one used here, the numbers are sufficient enough since usually the best fit is achieved after 10 to 15 generations. This is demonstrated on the right panel of Fig.~\ref{dvd}, where the development over time (i.e. generations) of each of the free parameters is shown for fit 7 (see Table~\ref{perres}). The run time for the best model was about 4 hours in total on a PC. 

The left side of this figure shows the artificial input spectrum with the best PIKAIA fit after 20 generations using a starting population of 30 trial solutions. In this test run, 4 parameters were determined by PIKAIA which are the dust temperature on the inner boundary $T_{dust}$, the abundances of 2 dust species $f_2$ and $f_3$ and the optical depth $\tau$. 

This test showed, on the one hand, that the program works as expected and on the other hand gave a first idea about how the different parameters interact with each other. Nevertheless, it has to be noted that the test was neither realistic concerning the complexity of typical observed spectra nor the number of generations and population members.  A more realistic approach will be presented in Sect.~\ref{iso}. 

Another point that needs to be mentioned is the CPU time needed for the calculations, which amounted to about two hours for the full procedure. The method presented in this paper might therefore not be able to compete with an experienced scientist doing a handfit.  Still it offers some advantages compared to the ``intuitive''  method, apart from not getting your hands dirty but letting the computer do your work. With the fitness value it offers a parameter which allows to compare your results in a quantitative way. Furthermore the process is much less biased than a fit done by hand and as such, again, much better to be used for general comparison.

\section{Application to ISO spectra of AGB stars}
\label{iso}
In order to test our routine with more realistic data, models for selected AGB stars, observed by the Infrared Space Observatory (ISO) \citep{isoref}, were calculated. The stars were chosen by following criteria: To start with, they had to be simple in a sense that their spectral appearance and thus mainly their dust composition is an easy (not more than three species) and more or less known and confirmed one.  Nevertheless, the chosen stars should represent a wide parameter range. Finally, another criterion was that there had already been some DUSTY model calculations done for that object, in order to compare our results with previous ones and thus to put the algorithm to another test.

\begin{table}[th]
\centering 
\caption{Basic properties of ISO test stars} 
\begin{tabular}{c c c c c} 
\hline\hline 
Source & ISO TDT number & Spectral class & Variability & Ref\\ 
\hline 
CE And & 80104817& M5 & Lb & 1\\
$o$ Cet & 45101201 & M7IIIe & Mira & 2\\
Z Cyg & 37400126 & M5e & Mira & 3\\
TY Dra & 74102309 & M8 & Lb & 1\\
S Pav & 14401702 & M7IIe & SRa & 1\\
SV Peg & 74500605 & M7 & SRb & 1\\
\hline
\end{tabular}
(1) \citet{olofsson02}; (2) \citet{loup93}; (3) \citet{groen99}

\label{isostars} 
\end{table}
All the data on the selected objects (see Table~\ref{isostars}) have been taken from the ISO archive and have been processed with ISO's automatic data-analysis pipeline called Off-Line Processing, v10.1. For further reduction and processing the ISO Spectroscopic Analysis Package (ISAP) was used \citep{sturm98}.  Since the major interest was on the overall shape and not on special features of the observed spectra, they have been rebinned to a resolution of $R=200$ in order not to waste processing time by an unnecessarily high number of data points.\footnote{The dust species examined in this paper all exhibit only such dust features which are broad enough to be resolved at R = 200. Thus no essential dust features are missed at this resolution.}
Before getting into detail with each single star, a remark has to be made concerning the 30~$\mu$m feature visible in all the spectra. This feature has not been taken into account in any of the calculations since it is very likely to be an instrumental artefact rather than a real feature. Although \citet{sloan03} suggest that it might not be completely artificial, there is still no consensus on that matter.

For fits done by hand this feature does not present any problem since it can be easily ignored during the fitting procedure by the scientist. On the contrary, PIKAIA treats every data point equally so fits were only done until 26~$\mu$m and thus the wavelengths above are not shown in the presented plots since they are not used in the analysis.

The dust composition of the individual stars has been taken from the literature. In the case of S Pav and CE And \citet{heras05} served as a source for the starting values of the calculations.  For  TY Dra, $o$ Cet  and Z Cyg the values were taken from \citet{poschPhD}. Finally, for \mbox{SV Peg} both sources have been used to determine the starting values. Although the compositions used presented a good starting point, they had to be slightly adapted in the run, especially those of \citet{heras05}. This is most probably due to the fact that for these tests only a simple black body was used as a DUSTY input while for them a composed spectrum of a black body, the observed photospheric ISO-SWS spectrum and a Rayleigh-Jeans extrapolation served as input. 

The results for each star are discussed in the following sections. Furthermore Table~\ref{tabu} gives an overview of all input and output values of the fits shown in Fig.~\ref{isotest} plus the assumed starting values used for the genetic algorithm. The first entry for each star gives the parameters for the fit done by \citet{poschPhD}. Thus there are no values for the population and generation entered. The second entry gives the start parameters for the PIKAIA runs. This is implicated by the values 0 for the generation and population and the missing fitness value. Finally the best result for each star is shown. CE And and S Pav do not have fits done by \citet{poschPhD} available. For SV Peg the results of 2 additional fits done with PIKAIA are given. Fig.~\ref{svpegdetail} shows those results in comparison to the other fits.

\begin{table*}[th]
\centering 
\caption{Summary of the input values and results of the fits shown in Fig.~\ref{isotest}.} 
\begin{tabular}{c c c c c c | c c c c c c c c c c}
\hline\hline 
&  &  &       Input    & & & & & & & Output & & & & &\\
\hline
Star & Pop. & Gen. & $T_{*}$ & $T_{dust}$ & $\tau_{10\mu{}m}$ & $f_1$ & $f_2$ & $f_3$ & $f_4$ & $f_5$ & $f_6$ & $f_7$ & $f_8$ & Fitness\\
& & &[K] & [K] &  & &  &  &  & & &  &  & &\\
\hline

CE And & 0   & 0       & 2700 & 500 & 0.19     &      &       &      & 24 & 72 & 4     &         & &\\
CE And & 90 & 120  & 2700 & 450 & 0.009   &      &       &      & 20 & 65 & 12  &        & 3 &3.9998\\
\hline
o Cet     &       &         & 2600 & 500 & 0.03      &      &100 &      &       &       &       &         & &6.3535\\
o Cet     & 0    & 0      & 2600 & 500 & 0.03      &     &  50 & 50 &       &       &       &         & &\\
o Cet     & 90  & 100 & 2200 & 650 & 0.2        &     &  60 & 40 &       &       &       &          & &6.3538\\
\hline
Z Cyg    &        &        & 2400 & 400 & 0.09      &      & 100 &      &       &       &       &      & &4.3745\\
Z Cyg    & 0    & 0      & 2500 & 500 & 0.15      &      &  50  & 50&        &       &       &        & &\\
Z Cyg    & 90  & 100 & 2700 & 490 & 0.13      &      &       &  30& 36 & 34 &       &        & &4.3926\\
\hline
TY Dra  &        &         & 3000 & 600 &0.025    &      &       &      &       &       &       &    100 & &9.2667\\
TY Dra  & 0    & 0      & 3000 &  600 &0.025   &      & 50  &50 &       &       &       &           & &\\
TY Dra  & 90  & 100 & 3000 & 650 & 0.3        &      & 60 & 30 &       & 10 &       &            & &9.6803\\
\hline
S Pav    & 0    & 0      & 2800 & 400 & 0.07      & 42 &      &      & 42 &        & 12 &           & 4 &\\
S Pav    & 100& 200 & 2800 & 650 & 0.0045 & 41 &      &      & 41 &        & 14 &           & 4 &9.9916\\
\hline
SV Peg &        &         & 3100 & 550& 0.008    &15  & 71&      &       &        &  10 &      & 4 & 6.3979\\
SV Peg & 0a  & 0a    & 3100 &  550& 0.008   &15  & 71&      &       &        &  10 &       & 4 & \\
SV Peg & 100 & 90  & 3000 & 615 & 0.008  &10  & 60&      &       &        &  26 &      & 4 &6.4002 \\
SV Peg & 100 & 120 & 3000 & 560 & 0.007  &13  & 61&      &       &        &   22 &       & 4 &6.4006 \\
SV Peg & 0b   & 0b    & 2800 & 500 &  0.12   &22  &      &      &  45 & 20 &   9  &       & 4 & \\
SV Peg & 100 & 200 & 2800 & 570 & 0.011  & 31 &      &      & 32 & 20  &  12&          & 5 &6.4017\\
\hline
\end{tabular}
\label{tabu} 
\end{table*}

\subsection{\object{CE And}}
The MIR spectrum of CE And is dominated by a classical amorphous silicate emission \citep{fabian01, heras05}. As a starting point for the dust composition amorphous  melilite, amorphous olivine and crystalline magnesiowustite, respectively $f_4$, $f_5$ and $f_6$,were chosen. During the first manual tests it appeared that there might exist another species contributing to the spectrum. In order to test this hypothesis crystalline spinel ($f_8$) was used as an additional species. It indeed turned out that the results where better with this additional component. Concerning the population number, a starting population of 90 individuals seemed to be sufficient for the given problem as for the number of generations 120 seemed to serve the purpose. The final results are shown hereafter in Table~\ref{tabu}.

\subsection{\object{$o$ Ceti}}
The reason for picking $o$ Ceti is obvious. As \textit{the} Mira star per se, its dust composition is very well known and it has been topic of a number of publications of which some presented synthetic fits \citep[e.g.][]{heras05}. Mira's mineralogy exhibits mostly silicate dust, which makes it a perfect target for the further evaluation of the code. 

As starting parameters for the calculation, we assumed the same dust composition as \citet{poschPhD}. After some iterations and efforts to adapt the parameters, it turned out that another dust composition including aluminum oxide produces a better fit. The best PIKAIA results  were achieved with 100 generations and a population of 90. Results are shown in Table~\ref{tabu} and also in Fig.~\ref{isotest}.

\subsection{\object{Z Cyg}}
Z Cyg is a very well known Mira variable with a solely silicate dust composition. The dust properties of Z Cyg have already been extensively discussed  by \citet{onaka02} focusing on the effects of time variations of SWS spectra. The SWS spectrum at variability phase $\phi$=0.97 (i.e. TDT37400126) used here is the same as  \citet{onaka02} used to derive the dust emissivity of Z Cyg.  The mass loss rate, obtained from their best fit, of 7~x~10$^{-8}$ (r$_*$/3~x~10$^{13}$~cm) seems to be in rather good  agreement with the one (4~x~10$^{-8}$ M$_{\odot}$/yr) obtained by \citet{young95}.

Figure~\ref{isotest} shows a typical fit done by hand (dashed line) taken from \citep{poschPhD} and the best PIKAIA fit (solid line) that could be accomplished so far. This fit exhibits an exclusively silicate containing dust composition. Although this fit covers the main silicate feature around 20~$\mu$m quite well, there seem to be some problems connected with the 9.7~$\mu$m feature. In comparison to the hand fit, the PIKAIA fit is covering this feature much better. A combination of  amorphous pyroxene, $f_{3}$, amorphous  melilite, $f_{4}$ and amorphous olivine, $f_{5}$ proved to produce the fittest results. Still this results are not completely satisfying yet.  Again, results are given in Table~\ref{tabu}.

\subsection{\object{TY Dra}}
\begin{figure*}
\centering
\includegraphics[bb=0 200 595 780,scale=0.9]{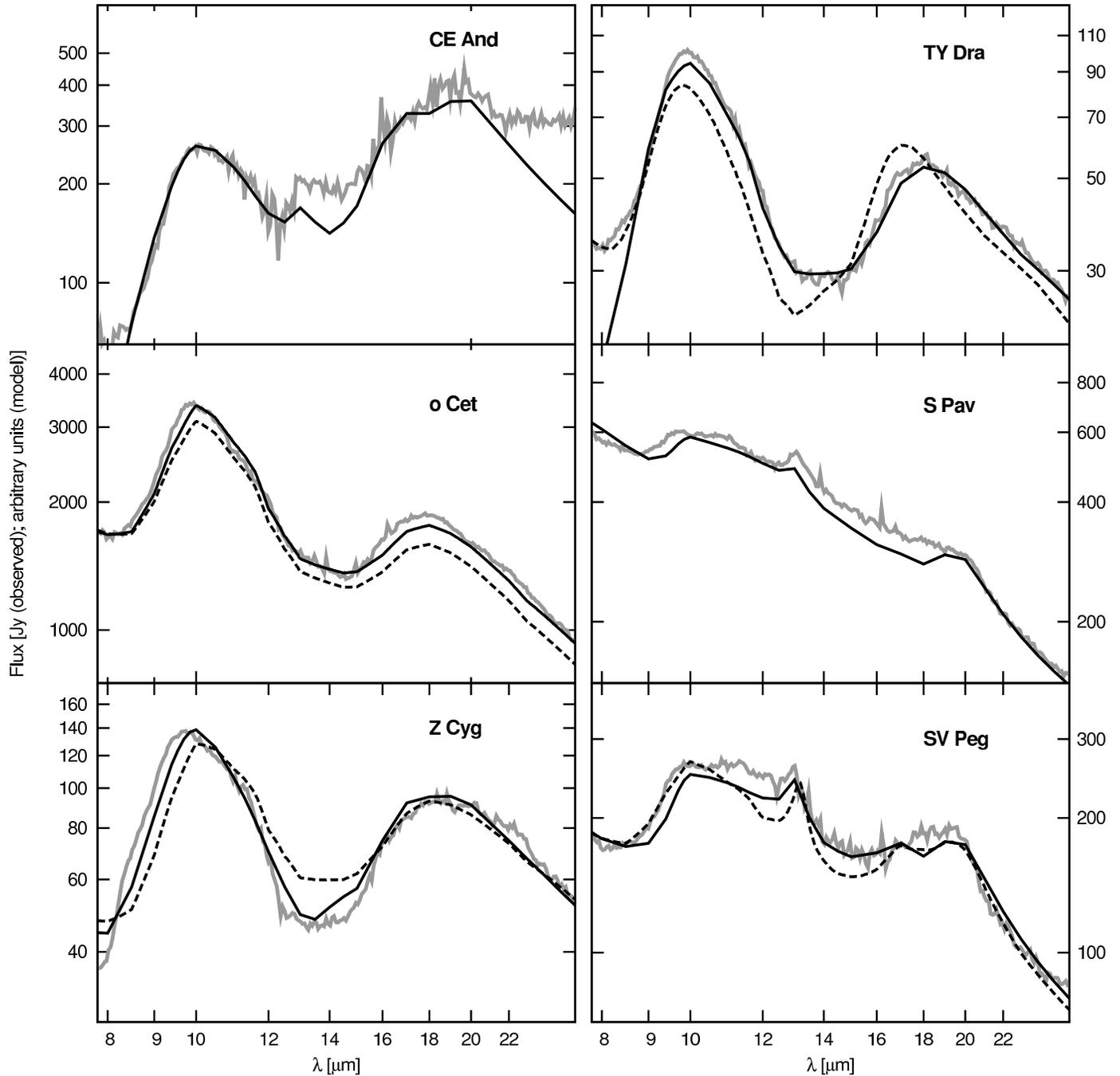}
\caption{This figure shows all the results of the PIKAIA fits in comparison to the fits done by \citet{poschPhD}, if available, for our test stars. The grey shaded solid line indicates the observed ISO spectrum of each star. The results of the Posch-fit are shown as dashed black line and finally the solid black line represents the best PIKAIA fit we could achieve so far. For detailed information about each star check the respective section and also Table~\ref{tabu} for the parameter values.}
\label{isotest}
\end{figure*}

TY Dra proved to be more complicated than first expected. Although its spectrum is supposed to be entirely dominated by amorphous silicate dust (MgFeSiO$_{4}$; see \citet{poschPhD}), this solution did not turn out to be the fittest for the given starting values. In Fig.~\ref{isotest} the hand fit based on the assumption mentioned is shown. It can be clearly seen that, even though the 10~$\mu$m peak seems to fit quite well, and with some adjustments of the input parameters might fit even better, the peak around 19~$\mu$m is not well represented at all. On the other hand, the PIKAIA fit seems to be a better option, although not perfect, for this case. Here we assumed a dust composition consisting of amorphous pyroxene, amorphous aluminum oxide and amorphous olivine (respectively $f_{2}$,$f_{1}$ and $f_{5}$). 

\subsection{\object{S Pav}}
S Pav is a fairly complex star in terms of  dust composition \citep[see e.g.][]{fabian01}. As stated in \citet{heras05}, the given starting composition of amorphous aluminium oxide, amorphous melilite, magnesiowustite and crystalline spinel (respectively $f_1$,$f_4$,$f_6$ and $f_8$) PIKAIA was able to produce a fairly reasonable fit for the given parameter range presenting only minor changes in the amount of amorphous aluminium oxide, amorphous melilite and magnesiowustite. Still, setting the generation number to 200 in order to obtain the best fit possible, went at the cost of run time. 
\subsection{SV Peg}
Just like S Pav, SV Peg shows a complex mineralogy. Although still dominated by silicates, its spectrum also exhibits quite strong features of amorphous  aluminium oxide, crystalline magnesiowustite and crystalline spinel according to \citet{poschPhD}. This complex dust composition made the fitting process for our routine very complicated.

SV Peg turned out to be a surprise in terms of fitting. Looking like too complex to be fitted by this DUSTY-PIKAIA routine in the beginning, the results were fairly good in the end, but at the costs of having to set very strict constraints to the input dust composition.

\begin{figure}
\centering
\includegraphics[bb=10 10 974 550,scale=0.35]{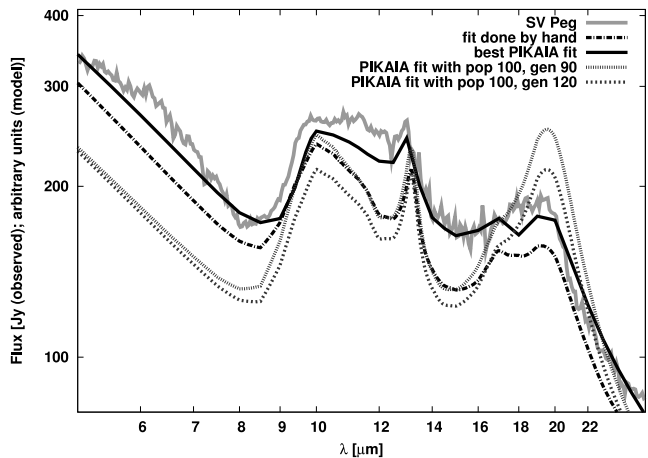}
\caption{This Figure shows a fit of the spectrum of SV Peg done by Posch (black, dashed dotted line) in comparison to 3 PIKAIA fits. The best of those fits (black line) shows already a quite good approximation of the given observed spectrum, while the other fits (small and big dotted grey) display a kind of evolution during the fitting process. It can be seen that with an increasing number of generations the fit approaches the observed spectrum. Still even with a higher number of generations the results were not satisfying, so we decided to change the dust settings of the model. This led us to the best fit, which can be also seen in Fig.~\ref{iso}. }
\label{svpegdetail}
\end{figure}

Figure \ref{svpegdetail} tries to give an impression about the evolution of fits and the difficulties that appear when choosing the input parameters. The two fits displayed in grey (further referred to as fit A and fit B) show how the increase of the generation number leads to an improvement of the fit. Nevertheless, it turned out that in the case of these two fits, the chosen dust parameters were not the most suitable ones, since a restriction in the parameter space and the raise of the generation number did not have such a strong effect on the results as expected.

Fit A was done with a dust composition of amorphous  aluminum oxide,amorphous silicate,  crystalline magnesiowustite and crystalline spinel (respectively   $f_1$,$f_2$,$f_6$ and $f_8$) suggested by  \citet{poschPhD}. This worked very well for the manual fit. Unfortunately it turned out that a too high number of free parameters combined with too large range of values for the respective parameters did not perform very well. This was mainly due to the chosen number of only 90 generations (and also population members). For a problem of this complexity more time is needed to find a solution, thus more generations are needed.
As a consequence the number of generations for fit B was increased to 120. It can be seen clearly in Fig.~\ref{svpegdetail}, that the fit  slowly improves and approaches the real spectrum. Nevertheless only raising the number of generations did not have such a strong effect as desired. Furthermore, the increment lead to an immensely extended run time of almost 2 days.

Since we were not able to achieve  considerably better results even with higher generation numbers, a slightly altered dust composition, consisting of marphous aluminium oxide, am. melilite, olivine, crystalline magnesiowustite and crystalline spinel (respectively   $f_1$,$f_4$,$f_5$,$f_6$ and $f_8$) in accordance to \citet{heras05}, was used. This time only the amount of dust was set to be a free parameter. All other variables were fixed. Furthermore the number of generations was increased to 200 to provide PIKAIA with enough time to arrive the best fit possible. This result can be also seen in Fig.~\ref{svpegdetail} as well as in Table~\ref{tabu}.

\section{Conclusions and Outlook}
AGB stars are known to exhibit strong dust features in the IR range of their spectra. The carriers of those features have to be identified in order to study the dust composition of these shells and hence the mass loss process of AGB stars. 

The goal of this work was to study an automatised routine to fit this IR part of the spectrum of an AGB star and to explore the possibilities and more important the boundaries for a routine like this. Therefor we have focused in this work only on the improvement of  fits which have already been carried out on the respective objects. 
This goal was definitely reached. We could show that it is indeed possible to improve fits done by various researchers using a genetic algorithm for fine tuning of the parameters. We could also show that in cases of Z Cyg, SV Peg and TY Dra a more elaborate or even slightly different dust composition leads to better results than with the composition taken from  \citet{poschPhD}, which was used as a starting point for our calculation. For CE And, o Cet, S Pav and SV Peg our results showed only minor difference from the dust compositions taken from \citet{heras05}, which we used as starting parameters for our models for these objects.

Furthermore this routine also provides a quantitative method, namely the fitness value supplied by the genetic algorithm, for comparison of the model spectra. This is connected with the need of independence from an all hands on method done by a scientist, which is, even when carefully trying to avoid it, always biased by the individual experience on the respective field of research. This method can be seen as a little step away form this type of fitting process to a more automatised and independent fitting process.

Nevertheless there are a few things which need to be kept in mind while using PIKAIA and DUSTY in order to fit a spectrum.
First, the starting parameters for the fits of all stars mentioned in this paper  were taken from the literature. However, one will not always work on objects that have already been the subject of investigations beforehand. This leaves us with the question  of how to determine the basic stellar parameters for each object needed as input for our automatic fitting method. For a star's temperature the most likely approach is carrying out JHK-photometry in order to retrieve these values. Another logical approach would be using only similar object of a certain group, which then should also offer some hints for the approximate values.

The dust temperature is also a value which might not be found  in the literature. In order to estimate the temperature a simple black body fit \citep[e.g.][] {kerschbaumPhD} of the respective wavelength region should offer a sufficient starting value for our calculations.

Still, the by far most difficult part in choosing the starting values for a model is the selection of suitable dust species and their respective amount. So far the only reliable method is an experienced scientist. Although this step could be executed by a program doing a comparison of local maxima in the dust spectra with the star's spectrum, an accordance in the peaks unfortunately does not necessarily mean that a single species or a certain combination applies to the star.
For now human intervention in this step seems to be unavoidable and a performance efficient  black box fitting procedure out of reach.

This leads also directly to the problem of the run time of our routine. \citet{metcalfe00} as well as \citet{puls05} used the parallelized version of PIKAIA to run their simulations. In this paper we used only a serial version of PIKAIA in order to study the overall behaviour of the routine in combination with DUSTY in a setting as simple as possible. Thus a full optimisation procedure took on an average work station up to a day at most. 

The most time during the fitting procedure is consumed by  DUSTY  and not by PIKAIA itself. Thus, our attempts to speed up the routine focused on DUSTY. We reduced the wavelength grid as much as possible, skipped unnecessary command line outputs and used the basic black bodies, inbuilt in DUSTY, as a background spectrum instead of a more elaborated one. Another problem which increases the runtime, is that DUSTY models with a good fitness, that continue to the next generation are not stored anywhere, but are calculated again.  All those points together will likely lead to an extension of the runtime by up to a factor of two. In the current version of our code we did take these points into account as we were aiming at testing the general fitting possibilities of this approach first. Solutions to most of these points are straightforward and their implementation is foreseen for a forthcoming version of our code currently under development. The results shown in this paper are very promising, thus we expect that our method, keeping in mind its limitations, will finally provide a valuable tool for fitting dust spectra of AGB stars.
\begin{acknowledgements}
A. B. received a DOC-fFORTE grant from the Austrian Academy of Sciences.  This work was supported by the \textit{Fonds zur F\"orderung der Wissenschaftlichen Forschung} (FWF) under project number P18939-N16.  TL acknowledges funding by the FWF under project number P20046-N16. Furthermore we would like to thank T. Posch for providing his fits to the ISO spectra and W. Nowotny for careful reading and helpful feedback.
\end{acknowledgements}
\bibliographystyle{aa} 
\bibliography{13968refs}

\end{document}